# Spin-Phonon Coupling, High Pressure Phase Transitions and Thermal Expansion of Multiferroic GaFeO$_3$: A Combined First Principles and Inelastic Neutron Scattering Study

M. K. Gupta[1], R. Mittal[1], M. Zbiri[2], Ripandeep Singh[1], S. Rols[2], H. Schober[2] and S. L. Chaplot[1]

[1]*Solid State Physics Division, Bhabha Atomic Research Centre, Mumbai 400085, India*
[2]*Institut Laue-Langevin, BP 156, 38042 Grenoble Cedex 9, France*

We have carried out an extensive phonon study on multiferroic GaFeO$_3$ to elucidate its dynamical behavior. Inelastic neutron scattering measurements are performed over a wide temperature range, 150 to 1198 K. First principles lattice dynamical calculations are done for the sake of the analysis and interpretation of the observations. The comparison of the phonon spectra from magnetic and non-magnetic calculations highlights pronounced differences. The energy range of the vibrational atomistic contributions of the Fe and O ions are found to differ significantly in the two calculation types. Therefore, magnetism induced by the active spin degrees of freedom of Fe cations plays a key role in stabilizing the structure and dynamics of GaFeO$_3$. Moreover, the computed enthalpy in various phases of GaFeO$_3$ is used to gain deeper insights into the high pressure phase stability of this material. Further, the volume dependence of the phonon spectra is used to determine its thermal expansion behavior.

## I. INTRODUCTION

Materials showing more than two ferroic properties (magnetism, electricity, and elasticity) simultaneously come under the umbrella of multiferroics whose characteristics include the emergence of simultaneous electric and magnetic orderings, offering therefore opportunities for multifunctional device applications. This justifies the intense research going on this class of materials, and the keen interest they are subject to, at both the fundamental and practical sides [1-14]. Magnetism in transition metals containing materials is induced by the active spin components in the d-shell levels. On the other hand, ferroelectricity occurs generally in the absence of d-electrons. Hence it is intriguing to observe multiferroicity since this phenomenon involves a simultaneous emergence of both the properties. Over the last few decades, various multiferroic materials have been discovered which exhibit magnetoelectric (ME) coupling. However, most of the magnetoelectric multiferroics possess magnetic and ferroelectric transition temperatures far from the ambient one. For example in the case of $BiMnO_3$ [14], the Curie temperature, $T_C$, is about 100 K while the Neel temperature, $T_N$, is close to 750 K. This results in a weak magnetoelctric coupling at the ambient conditions. Practically, the weak coupling materials are not potentially useful. However, there are few mechanisms allowing to tune these properties simultaneously; like magnetic ferroelectricity induced by frustrated magnetism, lone pair effect, charge-ordering, and local non-centrosymmetry. For instance, the charge ordering driven magnetic ferroelectricity is observed in a large number of rare earth oxides [11,12]. Understanding the mechanism of multiferrocity is of considerable importance for the design of new multiferroics at controllable conditions (temperature and pressure). Hence the electric and magnetic properties attributed to the dynamics of ions and electrons need to be studied and explored.

$GaFeO_3$ belongs to the class of multiferroic compounds and shows a ME coupling at low temperature. It does not contain lead or bismuth species, making it ecologically and biologically attractive. At room temperature the structure [15] is chiral orthorhombic ($Pc2_1n$), while its parental oxides $Fe_2O_3$ and $Ga_2O_3$ occur [16] in the rhombohedral and monoclinic phases, respectively. The orthorhombic structure of $GaFeO_3$ has eight formula units per unit-cell, with two different symmetry inequivalent sites of iron and gallium atoms; Fe1, Fe2, and Ga1 and Ga2, respectively. The tetrahedral sites are occupied by Ga1, while Ga2, Fe1 and Fe2 occupy all the octahedral sites (Fig. 1). The spontaneous electric polarization is found [17] to be along the b axis. Ideally, the magnetic structure of $GaFeO_3$ is expected to reflect an antiferromagnetic ordering, since the magnetic moments of Fe1 and Fe2 cations are antiparallel. However due to the observed disorder on the Fe and Ga sites [15], a ferri-magnetic transition is observed below 225 K, instead [15]. The magnetization axis was found to be along c-axis.

In a first principles study, Han and coworkers [18] suggest that distorted octahedra, $GaO_6$ and $FeO_6$, in $GaFeO_3$ lead to a noncentrosymmetric structure, which might be responsible for the electric polarization. The authors also showed that the site disorder involving the interchange of Fe and Ga2 sites is highly probable and consistent with the presence of the observed Fe disorder [15] with the Ga2 site. They indicate that the minimum of total energy is reached when adopting an antiferromagnetic spin configuration. However, anti site disorder of Fe and Ga atoms between octahedral Ga and Fe sites lead to a finite magnetic moment, and $GaFeO_3$ behaves like a ferrimagnet. It has been concluded that significant orbital magnetic moment of Fe ions is attributed to the local distortion of oxygen octahedra and the off centering of the iron atoms.

Interestingly, the unequal distribution of Fe spins in $GaFeO_3$ is due to the Ga-Fe disorder. This material is known to exhibit piezoelectricity and ferrimagnetism, with a Curie temperature, $T_C$, of about 225 K. This could be enhanced by a site disorder between Ga and Fe. It has been shown that the $T_C$ could be enhanced [19] to ~ 350 K by increasing the Fe content to about 40% ($Ga_{2-x}Fe_xO_3$ (x=1.40). The magnetic structure and magnetoelectric properties of $Ga_{2-x}Fe_xO_3$ (0<x<1.1) were extensively studied by T. Arima and coworkers [15]. The authors found that the saturated magnetization as well as the ferrimagnetic phase transition temperature increases with increase in Fe content, while the coercive force decreases. The linear and quadratic ME coefficients measurements show that the electric polarization is largely modulated when a magnetic field is applied parallel to the direction of the spontaneous magnetization. However it has a negligible effect when the field is applied parallel to the spontaneous polarization axis. Thin films of $GaFeO_3$ are reported to exhibit [20] ferroelectricity at room temperature, which makes them practically useful at the nano-level. We note also that the ball milling transforms [21] the structure of $GaFeO_3$ from orthorhombic to rhombohedral (R3c).

First Principle studies of zone centre phonon modes and Raman measurements were reported on the isostructural compound $AlFeO_3$ by Kumar and coworkers [22]. The Raman measurements have been performed in the temperature range 5 - 315 K. The observed spectra showed that the intensity of the Raman mode at 1230 cm$^{-1}$ vanishes to zero above 250 K. It was concluded that this mode originates from a two magnon Raman process. The authors also reported first principles calculation of the zone center phonon modes in magnetic ordered and disordered structure. They found a strong interaction between spin and lattice vibrations [22].

X-Ray as well as neutron diffraction, dielectric, Raman and IR measurements have been reported on GaFeO$_3$ [23-28]. No structural phase transition was observed [28] in the temperature range 14 - 1368 K. A dielectric anomaly [24] has been observed at the magnetic transition temperature. A spin-phonon coupling is reported [26] to take place below 210 K by observing the discontinuity in the peak position of the Raman mode at 374 cm$^{-1}$. Raman and Mossbauer spectroscopic studies on GaFeO$_3$ have also been reported [27]. The authors observed a disordered nature of the compound. The peak width of the phonon mode at 700 cm$^{-1}$ shows an anomalous large broadening around the Curie temperature, which is a measure of anharmonicity. The data was interpreted within the context of coupling of phonons and the Fe spins. Further, the stability of GaFeO$_3$ has been studied [29] under pressure; up to about 65 GPa. The compound undergoes a phase transition [29] from Pc2$_1$n to Pbnm phase at about 25 GPa. Increasing further the pressure to 53 GPa, the Pbnm phase also undergoes first order phase transition due to quenching of the Fe magnetic moment. Spin waves measurements have been reported by inelastic neutron scattering [30-32] in similar systems (TmFeO$_3$, ErFeO$_3$, YFeO3 and TbFeO$_3$). It comes out that an incommensurate phase was evidenced [32] in TbFeO$_3$, upon applying a magnetic field.

The various studies available on GaFeO$_3$ are based on structural and electronic considerations. A limited amount of work on phonon dynamics has been reported, but it was restricted to the zone centre phonon modes. Presently, we provide a detailed analysis of lattice dynamics and spin phonon coupling in GaFeO$_3$, where both the zone-centre and zone-boundary modes are covered. A better understanding of the dynamics governing the thermodynamical aspects of this promising multiferroic looks necessary for future fundamental and practical developments. In this context, we have measured the phonon density of states over a wide temperature range 150-1198 K. We have computed the phonon spectrum from first principles density functional theory to quantitatively explore the dynamics. The study is done in the ordered phase, by first considering the magnetic interactions and then neglecting them to better explore the possible interplay and effect of the spin degrees of freedom on the lattice dynamics [33, 34]. Further, the total energy and enthalpy is estimated in various phases to determine the relative phase stability of GaFeO$_3$. The equation of state has been calculated and compared with the available experimental data. Additionally, the volume thermal expansion has also been calculated as to have a better view on the thermodynamical picture of GaFeO$_3$.

Our paper is organized as follows: Section II and Section III provide details on the inelastic neutron scattering investigations and computational technicalities, respectively. The results are discussed in four parts in section IV (A-D): IV-A deals with the temperature dependence of phonon spectra, IV-B highlights the effect of magnetic ordering on the calculated phonon spectra, in IV-C we treat essentially total energy and free energy calculations for the phase diagram purpose, and the thermal expansion behavior is given in IV-D. Finally, conclusions are drawn in Section V.

## II. EXPERIMENTAL DETAILS

The inelastic neutron scattering measurements were carried out using the direct-geometry thermal neutron IN4C spectrometer at the Institut Laue Langevin (ILL), France. The spectrometer is based on the time-of-flight technique and is equipped with a large detector bank covering a wide scattering angle range of about $10^o$ to $110^o$. The polycrystalline sample of $GaFeO_3$ was prepared by solid state reaction method [15]. About 10 grams of polycrystalline sample of $GaFeO_3$ has been used for the measurements. The measurements were performed at several temperatures in the range 150-1198 K. The low temperature measurements were performed using a standard orange cryostat. For the high temperature range, the sample was put into a quartz tube insert and mounted into a furnace. The other end of the quartz tube was kept open in the air.

For these measurements we have used an incident neutron wavelength of 2.4 Å (14.2 meV), performing in the up-scattering mode (neutron energy gain). The momentum transfer, Q, extends up to 7 Å$^{-1}$. In the incoherent one-phonon approximation, the measured scattering function $S(Q,E)$, as observed in the neutron experiments, is related [35] to the phonon density of states $g^{(n)}(E)$ as follows:

$$g^{(n)}(E) = A < \frac{e^{2W_k(Q)}}{Q^2} \frac{E}{n(E,T) + \frac{1}{2} \pm \frac{1}{2}} S(Q,E) > \qquad (1)$$

$$g^n(E) = B \sum_k \{\frac{4\pi b_k^2}{m_k}\} g_k(E) \qquad (2)$$

where the + or − signs correspond to energy loss or gain of the neutrons respectively and where $n(E,T) = \exp(E/k_BT) - 1\ ^{-1}$. $A$ and $B$ are normalization constants and $b_k$, $m_k$, and $g_k(E)$ are, respectively, the neutron scattering length, mass, and partial density of states of the $k^{th}$ atom in the unit cell. The quantity between < > represents suitable average over all $Q$ values at a given energy. $2W(Q)$ is the Debye-Waller factor. The weighting factors $\frac{4\pi b_k^2}{m_k}$ for various atoms in the units of barns/amu are: Ga: 0.098; Fe: 0.208 and O: 0.265. The values of the neutron scattering lengths are taken from Ref. [36].

## III. COMPUTATIONAL DETAILS

Relaxed geometries and total energies were obtained using the projector-augmented wave (PAW) formalism [37,38] of the Kohn-Sham density functional theory [39, 40], within both the local density approximation (LDA) and the generalized gradient approximation (GGA), implemented in the Vienna *ab-initio* simulation package (VASP) [41]. The GGA was formulated by the Perdew–Burke–Ernzerhof (PBE) density functional [42]. The LDA was based on the Ceperly–Alder parametrization by Perdew and Zunger [43]. Both non-spin-polarized and spin polarized calculations were performed. The magnetic calculations have been carried out for the A-type antiferromagnetic ordering in the $Pc2_1n$ phase. Moreover, since $GaFeO_3$ is known to be a Mott insulator, the on-site Hubbard correction is applied within the Dudarev approach [44] using $U_{eff}$=4 eV [45-49]. All results are converged well with respect to *k*-mesh and energy cutoff for the plane wave expansion. The break conditions for the self-consistent field (SCF) and for the ionic relaxation loops were set to $10^{-8}$ eV and $10^{-5}$ eV Å$^{-1}$, respectively. The latter break condition means that the obtained Hellmann–Feynman forces are less than $10^{-5}$ eV Å$^{-1}$.

Both full (lattice constants and atomic positions) and partial (only atomic positions) geometry relaxations were carried out. Hereafter, the labeling "FM" and "FNM" refer to fully relaxed magnetic and fully relaxed non-magnetic calculations. Further, "PNM" refers to the partially relaxed non-magnetic calculation, where we used the structure obtained from "FM" and relaxed only the atomic positions without magnetic ordering. The structural details relevant to the present calculations are summarized in Table I. Total energies were calculated for 60 generated structures resulting from individual displacements of the symmetry inequivalent atoms in the orthorhombic ($Pc2_1n$) phase, along the six inequivalent Cartesian directions ($\pm x$, $\pm y$ and $\pm z$). Phonons are extracted from subsequent calculations using the direct method as implemented in the Phonon software [50]. The free energy calculations of $GaFeO_3$ are also done in the Pbnm and R3c phases.

## IV. RESULTS AND DISCUSSION

### A. Temperature dependence of phonon spectra

The phonon spectra of $GaFeO_3$ (Fig. 2) were measured up to 1198 K, across the magnetic transition (~ 225 K). The magnetic signal is expected to be more pronounced at low Q, and to vanish at high Q, following the magnetic form factor. Therefore, two Q-domains were considered; i.e., high-Q (4 to 7 Å$^{-1}$) and low-Q (1 to 4 Å$^{-1}$).

The temperature dependence of the Bose factor corrected S(Q,E) plots of $GaFeO_3$ are shown in Fig 2. At low temperatures (upto 315 K), the low-Q data shows a larger elastic line as compared to the high-Q spectra. Presently, given the lack of detailed magnetic measurements, we speculate that this quasi-elastic scattering may originates from spin fluctuations which disappear at high temperatures. In the high temperature range, only phonons contributes significantly to the spectra, and therefore the width of the elastic line is similar in both the Q ranges.

The phonon spectra inferred from the S(Q,E) data, within the incoherent approximation, are also shown in Fig. 2. The phonon spectra consist of several peaks located around 20, 30, 55 and 80 meV. We find that both the high Q as well as the low Q data show large variation in the intensity as a function of temperature. At low energy (below 40 meV), the low Q data are more intense in comparison to the high Q data. Further for the low Q part, at 150 K below the magnetic transition temperature (~225 K), there is a large intensity of the low energy inelastic spectra (~ 20 meV) as compared to the data collected at higher temperatures. This is expected to be due to a strong magnetic signal. At 848 K, it is found that in both the low Q as well the high Q data, there is a considerable decrease of the intensity of the low energy peaks around 20 meV. Although $GaFeO_3$ undergoes a paramagnetic to ferri-magnetic transition [15] around 225 K, a paramagnetic scattering persists in the low energy range around 20 meV, at 240 and 315 K. The intensity in the higher energy range, above 55 meV, of the high Q data does not show significant temperature dependence, confirming a pure phonon contribution in these spectral regime. Above 848 K, there is a loss of intensity, due to paramagnetic scattering, and only phonons contribute in this range

$GaFeO_3$ does not show any structural phase transition at high temperature. However polyhedral ($GaO_4$, $GaO_6$, $FeO_6$) distortions are found to increase upon heating up to 1198 K [28]. This might be an additional reason for the broadening of the phonon spectra above 60 meV at high temperatures, besides the increased thermal amplitudes.

**B. Magnetic ordering and calculated phonon spectra**

The microscopic origin of the polarization in multiferroic materials is attributed to the hybridization of the electronic orbitals producing a polar charge distribution and ionic displacements from the related centro-symmetric positions. Hence, it is important to study the lattice dynamics in order to understand the ME properties of multiferroics. Detailed electronic structure calculations of $GaFeO_3$ are reported in the literature [18]. However, phonon studies over the whole Brillouin zone are missing. Calculations of (electronic) structure and dynamics would help to gain newer and deeper insights into the various physical properties and possible phase transitions of this kind of materials.

The calculated Fe magnetic moment in the equilibrium structure in the $Pc2_1n$ phase at Fe1 and Fe2 sites are 4.1 $\mu_B$ and 4.1 $\mu_B$, respectively, which is in agreement with the reported experimental values [15] of 3.9 $\mu_B$, 4.5 $\mu_B$. Neglecting the spin degrees of freedom in the calculations leads to a collapse of the b-lattice parameter, with a value decreasing from 9.29 Å to 8.77 Å. However by considering Fe magnetism, the calculated value of b-lattice parameter is brought to agreement with the observation (Table I).

Fig. 3 compares the experimental and calculated phonon spectra. The "FNM" calculation results in a shift of all the modes to higher energies. This is due to the fact that the b-axis is underestimated in FNM calculations, leading to an overestimation of the phonon energies. Interestingly, the model calculations "FM" and "PNM" provide a very good agreement with the experimental spectra. We notice however some differences in the low energy part of the phonon spectra. The difference comes in fact from the value of the Fe magnetic moment in the two numerical models. The main effect of the Fe spin degrees of freedom is to soften the calculated phonon energies around 30 meV, bringing them hence closer to the experimental values. This demonstrates the role of magnetic interactions in $GaFeO_3$, in a similar way to other recent phonon studies in other systems [33,34].

The "FM"-based calculated phonon spectra (Fig. 3) lead to peaks centered around 20, 30, 55 and 80 meV. The experimental spectra show peaks at 20 and 30 meV and clear humps at 55 and 80 meV. $GaFeO_3$ is known to have a Ga-Fe disorder, from diffraction measurements [15]. However our phonon calculations were done in the ordered phase (Table I). The structural disorder could lead to a large variation of the Ga/Fe-O bonds, and would then result in a broadening of the peaks, as experimentally observed.

The difference in the phonon spectra (Fig. 3) from the various calculations can be understood from the estimated atomistic contributions in terms of the partial density of states from LDA calculations (Fig. 4). The difference is primarily due to the nature of the chemical bonding, in the magnetic and nonmagnetic configurations, as well as the related volume effect. We find that vibrations of Fe and Ga atoms extend up to 45 meV, while the dynamics of the oxygen atoms spreads over the entire spectral range, up to 100 meV. The vibrational aspects due to the two Ga symmetry inequivalent atomic sites remain nearly invariant in all the three calculation types, while the Fe vibrations show a considerable change. The intensity of vibrational density of states of the Fe2 atoms is enhanced around 20 meV. The vibrations of Fe1 as calculated around 30 meV in the non-magnetic calculations are found to soften magnetically, and exhibit a peak around 20 meV. "FNM" calculations predict the oxygen vibrations to extend up to about 100 meV. The overestimation in the range of vibrations is primarily due to the non-inclusion of the Fe magnetic moment which results in a contraction of the unit cell. The "FM" and "PNM" model calculations show that the vibrations of all the oxygen atoms soften in the energy range 60 - 100 meV.. A further interesting finding consists of the vibrations of the O5 atoms, as extracted from the "FM" calculation type. The O5 atoms are connected only to the Fe1 and Fe2 atoms (Figure 1). The O5 vibrations (Figure 4) around 60 meV are related to the Fe magnetism. This dynamics is found to shift to lower energies at about 30 meV in the "FM" calculations.

Given the known effect of the density functional approximation (LDA or GGA) on the volume description (LDA tends to underestimate the volume value and GGA shows the opposite trend), we compare the "FM" calculated phonon spectra from LDA and GGA approaches. The unit cell volume from LDA and GGA calculations is estimated to be 405.3 Å$^3$ and 434 Å$^3$, respectively. The experimental value is 413.9 Å$^3$ [15]. The low energy part of the phonon spectra, which is sensitive to Fe magnetism, is nearly the same in both LDA and GGA (Figures 3 and 5). Above 50 meV, some variations are however noticed. The GGA calculated phonons above 50 meV are found to be slightly at lower energies as compared to LDA calculated phonons. Both the exchange-correlation methods lead to an overall good matching with the observations.

Under the orthorhombic (Pc2$_1$n) symmetry, GaFeO$_3$ possesses 240 zone centre modes corresponding to the irreducible representations: $\Gamma = 60A_1+60A_2+60B_1+60B_2$. Figure 6 compares the determined zone centre phonon modes from the various calculation types. The LDA and GGA approximations lead basically to the same phonon energies. Several modes are found to significantly differ when comparing the magnetic and non magnetic cases. This confirms a spin-phonon coupling behavior. The change in energies of the modes below 25 meV is mainly due to the magnetic interactions, while the high energy phonons are influenced by the volume effect.

**C. High pressure phase stability of GaFeO$_3$**

The high pressure measurements [29], up to 70 GPa (increasing and decreasing cycles), revealed a very rich phase diagram of GaFeO$_3$. Arielly and coworkers reported the emergence of a new orthorhombic phase (space group Pbnm) above 25 GPa, upon increasing pressure. The transition was found to fully establish at 45 GPa. In this phase all the Ga atoms have eight co-ordinations. However in the Pc2$_1$n phase, two different Ga sites are distinguishable; one with a six-fold symmetry, and the other possessing a four-fold coordination. Increasing further the pressure to about 53 GPa results in another first order transition with significant drop of the volume. However, the system remains in the same orthorhombic space group (Pbnm). At this pressure value (53 GPa), the magnetic interactions weaken due to the broadening of the iron d-bands. Mossbauer measurement reveals that the Neel temperature is close to 5 K, at about 77 GPa. Further decreasing the pressure to the ambient value, the hexagonal R3c phase was found to be the stable one, which is different from the originally starting orthorhombic Pc2$_1$n phase, at ambient conditions.

In the literature [29], only the lattice parameters of GaFeO$_3$ are available in the Pbnm and R3c phases. The related atomic co-ordinates are missing. We have therefore started from the atomic co-ordinates of LuFeO$_3$ and LiNbO$_3$, as provided in Refs. [52] and [53] respectively. Mossbauer spectroscopy reveals the existence of magnetic ordering in GaFeO$_3$ [29] even at high pressures. The crystal structure of GaFeO$_3$ in Pbnm and R3c phases has been calculated by relaxing the atomic co-ordinates as well as lattice parameters. The total energy has been calculated in both the phases in various antiferromagnetic configurations represented by the A, C, and G ordering types. Computationally, we found that the Pbnm phase is the most stable when adopting the G-type antiferromagnetic ordering, while the R3c phase stabilizes with the A-type antiferromagnetism. The calculated structural details under the Pbnm and R3c phases at 25 GPa and ambient pressure, respectively, are given in Table II. Therein the available experimental lattice parameters are also shown.

Presently, the high-pressure equation of state, total energy ($\Phi$) and enthalpy (H=$\Phi$+PV) of various phases of GaFeO$_3$ were estimated for the fully relaxed magnetic (FM) configuration. The GGA calculated enthalpy showed that the high-pressure Pbnm phase is more stable than the Pc2$_1$n phase at ambient pressure. Fig. 7(a) presents the enthalpy difference from LDA calculations, for the Pbnm and R3c phases with respect to the $Pc2_1n$ phase. Above 23 GPa, the Pbnm phase is found to be stable when comparing to $Pc2_1n$. The application of pressure leads to a change in the correlation between the electronic motions and affects the magnetic interaction. A quenching of the Fe magnetic moment in the Pbnm phase is found at 47 GPa, which triggers a sudden drop of the volume and increases the total energy (Fig. 7(b)). This is in agreement with the high pressure data [29] which shows a similar behaviour around 53 GPa. The values of the magnetic momemts on the Fe-atoms remain about 4.1 $\mu_B$ from ambient pressure to below 47 GPa and then decrease to 1.0 $\mu_B$ at this transition.

The experimental data showed [29] that on release of the pressure from 65 GPa, the R3c phase is stabilized from 53 GPa and down to ambient conditions. The ball milling is also known to transform [21] the structure of $GaFeO_3$ from orthorhombic ($Pc2_1n$) to rhombohedral (R3c). The calculated enthalpy under the $Pc2_1n$ phase is -7.196 eV/atom, while in the R3c phase the value is -7.209 eV/atom. This indicates that the R3c phase is more stable as compared to $Pc2_1n$. The calculated energy difference between the two phases is rather small (~13 meV/atom). It should be noted that the enthalpy of $GaFeO_3$ has been calculated for ordered structure of $GaFeO_3$ (Tables I and II). However, the experimental data indicate a disorder [15] on the Fe and Ga sites, which may lead to stabilization of $GaFeO_3$ in the $Pc2_1n$ phase at ambient pressure. The application of external pressure by ball milling or by diamond anvil cell may be responsible for the tuning of the small difference in the enthalpy, which would result in stabilizing the R3c phase of $GaFeO_3$.

The calculated phase diagram is qualitatively in a good agreement with the observation. It should be mentioned that it is difficult to identify experimentally the high pressure equilibrium phases, due to the large hysteresis. Fig. 8 shows the comparison between the LDA-calculated and experimental relative change of the unit cell volume in various phases of $GaFeO_3$ as a function of pressure. A very good agreement is noticed between our calculations and the measurements [29] in the $Pc2_1n$ and R3c structures; however, the volume in the Pbnm phase is underestimated. Table III gathers the LDA and GGA calculated elastic constants. The estimated bulk modulus values from LDA and GGA calculations, in the $Pc2_1n$ phase, are 207 and 178 GPa, respectively. The LDA determined value is found to be in a better agreement with the experimental bulk modulus value (226 GPa) [29]. As expected, the GGA underestimates the elastic constants by about 15% with respect to LDA, given that GGA tends to overestimate the calculated unit cell volume. This results in lowering the calculated bulk modulus values.

**D. Thermal expansion Behavior**

The thermal expansion behavior of any material is of considerable importance, since it plays a key role for potential applications. The calculation of the thermal expansion of $GaFeO_3$ is carried out within the quasi-harmonic approximation (QHA). In QHA, each phonon mode contributes to the volume thermal expansion coefficient [53, 54], given by: $\alpha_V = \frac{1}{BV} \sum_i \Gamma_i C_{Vi}(T)$, with $\Gamma_i (= -\partial ln E_i / \partial ln V)$ and $C_{vi}$ are the mode Grüneisen parameter and the specific heat of the $i^{th}$ vibrational state of the crystal, respectively. The volume dependence of phonon modes is calculated in the entire Brillouin zone. The pressure dependence of the phonon spectra in the entire Brillouin zone was extracted from LDA and GGA "FM" calculations, at two pressure points: ambient and 0.5 GPa. Figure 9(a) shows the calculated Grüneisen parameter values, $\Gamma(E)$. They show considerable variation as a function of the energy and are found to lie within 0.2-4.0. The thermal expansion behaviour has been calculated up to 1500 K. Neutron diffraction measurements on $GaFeO_3$ reported the absence of any high-temperature structural phase transition (up to 1368 K) [28]. The comparison between the experimental and calculated thermal expansion character is presented in Figure 9(b). GGA leads to a better agreement with the experimental data, while the LDA was found to underestimate the thermal expansion behavior.

**V. CONCLUSIONS**

We have reported the calculated and measured phonon spectra of the multiferroic material $GaFeO_3$. The measurements were performed over a wide temperature range (150-1198 K), covering all the relevant characteristic transition temperatures. Across the magnetic transition temperature, at 225 K, there is an increase of the intensity of the low energy phonons around 20 meV, associated with the dynamics of the Fe atoms. The low energy vibrations exhibit a significant Q dependence up to about 848 K, indicating a persistence of the paramagnetic spin fluctuations up to very high temperatures. $GaFeO_3$ is not subject to any structural high-temperature phase transition. However, the increase of the distortion amplitudes of the various polyhedral units might be at the origin of the gradual broadening of the stretching modes around 60 meV. The ab-initio phonon calculations highlighted unambiguously a spin-phonon coupling in $GaFeO_3$. The enthalpy calculations in various phases showed that the quenching of the Fe magnetic moment leads to the observed high pressure structural phase transition at 47 GPa. The calculated thermal expansion is in good agreement with the available experimental data.

TABLE I. Comparison between the experimental (4 K) and calculated (0 K) structural parameters of GaFeO$_3$ (orthorhombic phase, space group Pc2$_1$n). The experimental structure (space group Pc2$_1$n) consists [15] of all the atoms at 4$a$($x$, $y$, $z$). Wyckoff site with site occupancies of 1.0, 1.0, 1.0, 1.0, 1.0, 1.0, 0.82/0.18, 0.65/0.35, 0.23/0.77 and 0.30/0.70 of O1, O2, O3, O4, O5, O6, Ga1(Ga/Fe), Ga2(Ga/Fe), Fe1(Ga/Fe), Fe2(Ga/Fe) atomic sites, respectively. The ab-initio calculations were performed adopting and integer site occupancy. "FM", "FNM" and "PNM" refer to fully relaxed magnetic, fully relaxed non-magnetic and partially relaxed non magnetic calculations, respectively.

|     |         | Expt.  | FM(GGA) | FM(LDA) | PNM(LDA) | FNM(LDA) |
|-----|---------|--------|---------|---------|----------|----------|
|     | $a$ (Å) | 8.7193 | 8.8516  | 8.6610  | 8.6610   | 8.4791   |
|     | $b$ (Å) | 9.3684 | 9.5232  | 9.2923  | 9.2923   | 8.7713   |
|     | $c$ (Å) | 5.0672 | 5.1491  | 5.0355  | 5.0355   | 4.9999   |
| O1  | $x$     | 0.3228 | 0.3221  | 0.3233  | 0.3154   | 0.3255   |
|     | $y$     | 0.4262 | 0.4268  | 0.4291  | 0.4405   | 0.4517   |
|     | $z$     | 0.9716 | 0.9825  | 0.9836  | 0.9860   | 0.9802   |
| O2  | $x$     | 0.4864 | 0.4868  | 0.4857  | 0.4853   | 0.4789   |
|     | $y$     | 0.4311 | 0.4323  | 0.4330  | 0.4413   | 0.4555   |
|     | $z$     | 0.5142 | 0.5190  | 0.5190  | 0.5331   | 0.5312   |
| O3  | $x$     | 0.9979 | 0.9970  | 0.9969  | 0.9877   | 0.9851   |
|     | $y$     | 0.2022 | 0.2022  | 0.2014  | 0.2091   | 0.2216   |
|     | $z$     | 0.6541 | 0.6579  | 0.6564  | 0.6599   | 0.6605   |
| O4  | $x$     | 0.1593 | 0.1615  | 0.1621  | 0.1564   | 0.1590   |
|     | $y$     | 0.1974 | 0.1996  | 0.2005  | 0.2049   | 0.2123   |
|     | $z$     | 0.1480 | 0.1570  | 0.1575  | 0.1684   | 0.1662   |
| O5  | $x$     | 0.1695 | 0.1683  | 0.1677  | 0.1667   | 0.1651   |
|     | $y$     | 0.6717 | 0.6726  | 0.6742  | 0.6820   | 0.7001   |
|     | $z$     | 0.8437 | 0.8422  | 0.8447  | 0.8245   | 0.8309   |
| O6  | $x$     | 0.1736 | 0.1671  | 0.1664  | 0.1658   | 0.1632   |
|     | $y$     | 0.9383 | 0.9391  | 0.9394  | 0.9365   | 0.9509   |
|     | $z$     | 0.5166 | 0.5180  | 0.5217  | 0.5247   | 0.5372   |
| Fe1 | $x$     | 0.1538 | 0.1539  | 0.1549  | 0.1678   | 0.1709   |
|     | $y$     | 0.5831 | 0.5834  | 0.5836  | 0.5894   | 0.6049   |
|     | $z$     | 0.1886 | 0.1857  | 0.1883  | 0.1691   | 0.1689   |
| Fe2 | $x$     | 0.0346 | 0.0316  | 0.0308  | 0.0269   | 0.0186   |
|     | $y$     | 0.7998 | 0.7956  | 0.7961  | 0.8000   | 0.8168   |
|     | $x$     | 0.6795 | 0.6721  | 0.6739  | 0.6772   | 0.6785   |
| Ga1 | $x$     | 0.1500 | 0.1520  | 0.1510  | 0.1503   | 0.1462   |
|     | $y$     | 0.0    | 0.0     | 0.0000  | 0.0000   | 0.0000   |
|     | $z$     | 0.1781 | 0.1749  | 0.1770  | 0.1789   | 0.1873   |
| Ga2 | $x$     | 0.1593 | 0.1608  | 0.1607  | 0.1561   | 0.1589   |
|     | $y$     | 0.3073 | 0.3089  | 0.3095  | 0.3139   | 0.3204   |
|     | $z$     | 0.8106 | 0.8167  | 0.8160  | 0.8189   | 0.8181   |

TABLE II. The calculated structural parameter of GaFeO$_3$ in the orthorhombic (Pbnm) and hexagonal (R3c) phases within the local density approximation (LDA) in the fully relaxed magnetic structure. In the orthorhombic phase the O1, O2, Fe and Ga atoms are located at *4c* (*x*, *1/4*, *z*), *8d* (*x, y, z*), *4b*(*1/2*, *0*, *0*) and *4c* (*x*, *1/4*, *z*), respectively, while in the hexagonal phase O, Fe and Fe occupy the positions *36f* (*x*, *y*, *z*), *12c* (0, 0, *z*) and *12c* (0, 0, *z*), respectively. The experimental lattice parameters are from Ref. [29].

|     |   | Orthorhombic Pbnm phase | |
|-----|---|-------------------------|----------------|
|     |   | Expt. (25.7 GPa)        | Calc. (25 GPa) |
|     | *a* | 4.948(4)              | 4.793          |
|     | *b* | 5.165(20)             | 4.965          |
|     | *c* | 7.0000(8)             | 7.241          |
| O1  | *x* |                       | 0.413          |
|     | *y* |                       | 0.250          |
|     | *z* |                       | 0.142          |
| O2  | *x* |                       | 0.323          |
|     | *y* |                       | 0.076          |
|     | *z* |                       | 0.672          |
| Fe  | *x* |                       | 0.500          |
|     | *y* |                       | 0.000          |
|     | *z* |                       | 0.000          |
| Ga  | *x* |                       | 0.059          |
|     | *y* |                       | 0.250          |
|     | *z* |                       | 0.987          |
|     |   | Hexagonal R3c phase     | |
|     |   | Expt. (0.2 GPa)         | Calc. (0 GPa)  |
|     | *a* | 5.036(2)              | 4.981          |
|     | *b* | 5.036(2)              | 4.981          |
|     | *c* | 13.585(7)             | 13.425         |
| O   | *x* |                       | 0.969          |
|     | *y* |                       | 0.333          |
|     | *z* |                       | 0.080          |
| Fe  | *x* |                       | 0.000          |
|     | *y* |                       | 0.000          |
|     | *z* |                       | 0.018          |
| Ga  | *x* |                       | 0.000          |
|     | *y* |                       | 0.000          |
|     | *z* |                       | 0.309          |

TABLE III. The calculated elastic constants (in GPa units) of $GaFeO_3$ in the orthorhombic phase (space group $Pc2_1n$) in the fully relaxed magnetic structure at ambient pressure.

| Elastic Constant | GGA | LDA |
|---|---|---|
| $C_{11}$ | 291.8 | 344.6 |
| $C_{12}$ | 137.2 | 163.3 |
| $C_{13}$ | 119.8 | 148.4 |
| $C_{22}$ | 257.5 | 300.0 |
| $C_{23}$ | 127.0 | 159.0 |
| $C_{33}$ | 250.3 | 284.6 |
| $C_{44}$ | 62.5 | 72.7 |
| $C_{66}$ | 83.7 | 95.1 |

FIG. 1 (Color online) Crystal structure of $GaFeO_3$ in the $Pc2_1n$ space group. The atoms are labeled following Table I.

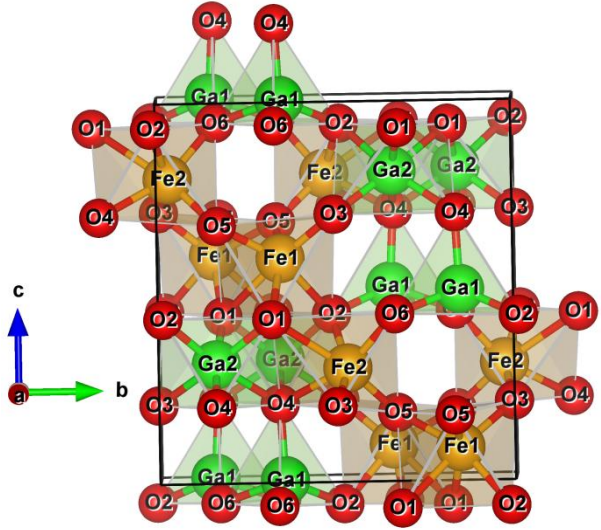

Fig. 2 (Color online) Temperature dependent inelastic neutron spectra of $GaFeO_3$. Top panel: the low-Q and high-Q Bose factor corrected S(Q,E), where both the energy loss (0 - 10 meV) and the energy gain (-100 - 0 meV) sides are shown. Bottom panel: the low-Q and high-Q, unity-normalized, phonon density of states, $g^{(n)}(E)$, inferred from the neutron energy gain mode S(Q,E) data, within the incoherent approximation.

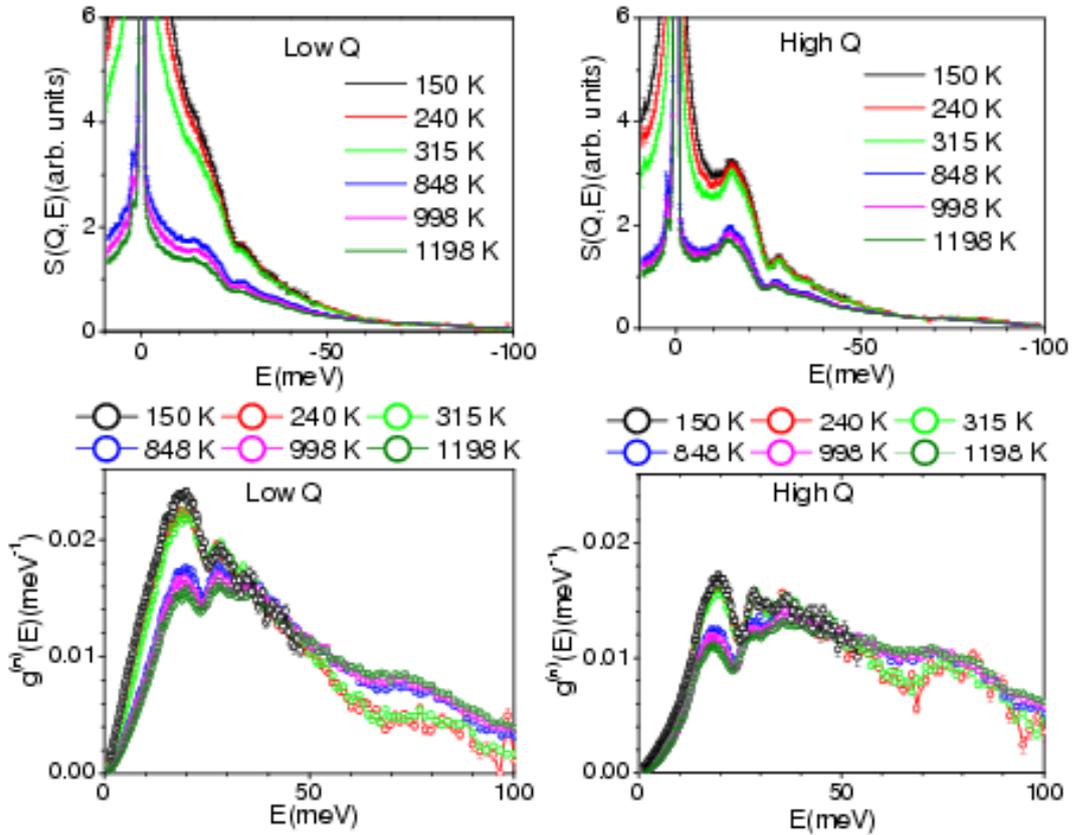

FIG. 3 (Color online) The calculated and experimental neutron inelastic scattering spectra of GaFeO$_3$. The experimental data consist of the "High Q" data collected at 315 K. The calculated spectra have been convoluted with a Gaussian of FWHM of 15% of the energy transfer in order to describe the effect of energy resolution in the experiment. All the spectra are normalized to unity in the entire spectral range. "FM", "FNM" and "PNM" refer to fully relaxed magnetic, fully relaxed non-magnetic and partially relaxed non magnetic calculations, respectively. For better visibility, the experimental and calculated phonon spectra are shifted vertically with respect to each other.

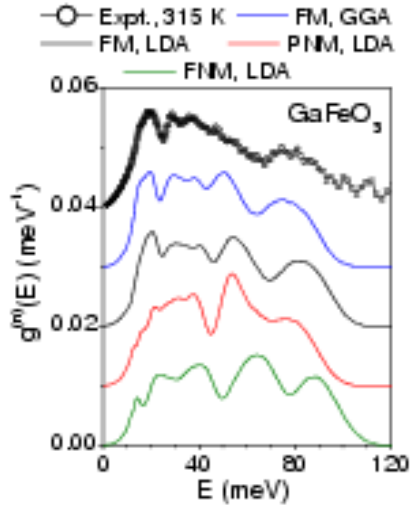

FIG. 4 (Color online) The calculated partial phonon density of states of various atoms in GaFeO$_3$ within the local density approximation (LDA). The atoms are labeled following Table I. "FM", "FNM" and "PNM" refer to fully relaxed magnetic, fully relaxed non-magnetic and partially relaxed non magnetic calculations, respectively.

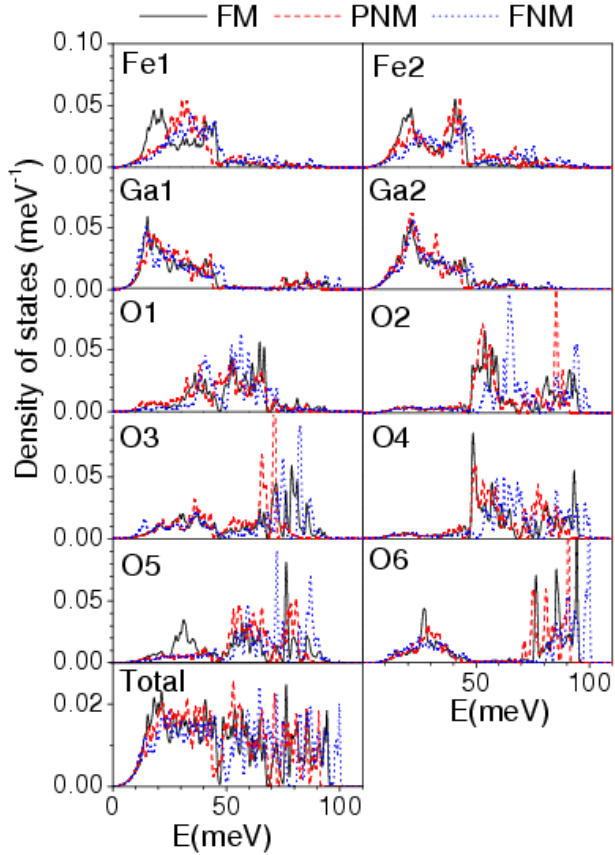

FIG. 5 (Color online) The calculated partial phonon density of states of various atoms in GaFeO$_3$ within the local density approximation (LDA) and the generalized gradient approximation (GGA) in the fully relaxed magnetic (FM) structure in Pc2$_1$n space group. The atoms are labeled following Table I.

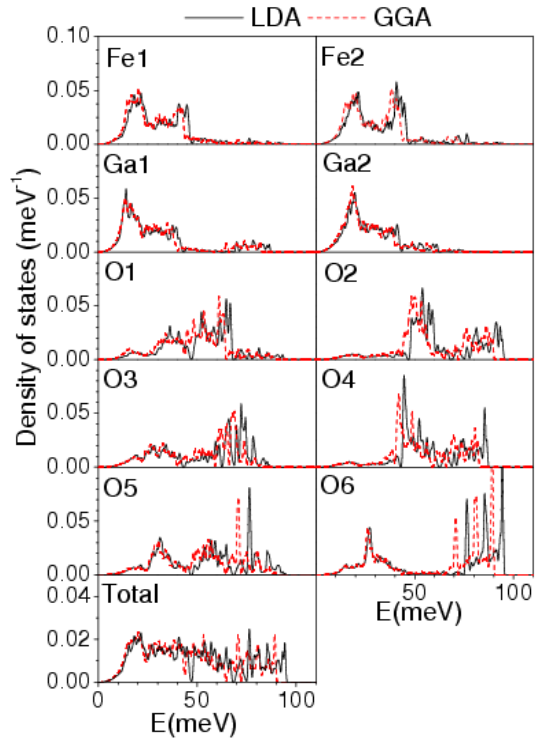

FIG. 6 (Color online) The calculated zone centre phonon modes of GaFeO$_3$ (orthorhombic phase, space group Pc2$_1$n). "FM", "FNM" and "PNM" refer to fully relaxed magnetic, fully relaxed non-magnetic and partially relaxed non magnetic calculations, respectively. Open and closed symbols correspond to calculations performed within the local density approximation (LDA) and generalized gradient approximation (GGA), respectively. A1, A2, B1 and B2 correspond to the group theoretical representations of the system symmetry.

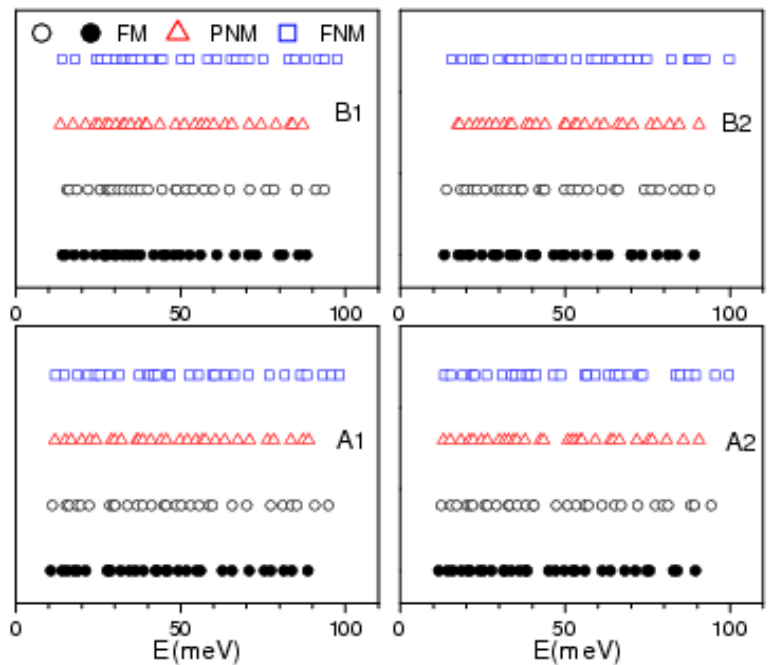

FIG. 7 (Color online) (a) The calculated enthalpy (H=Φ+PV) difference in the Pc2$_1$n and R3c phases with respect to the Pbnm phase of GaFeO$_3$ as a function of pressure within the local density approximation (LDA). (b) The calculated total energy (Φ) in the Pbnm phase of GaFeO$_3$ as a function of pressure within the LDA. The quenching of the Fe magnetic moment at 47 GPa leads to an increase of the total energy in the Pbnm phase.

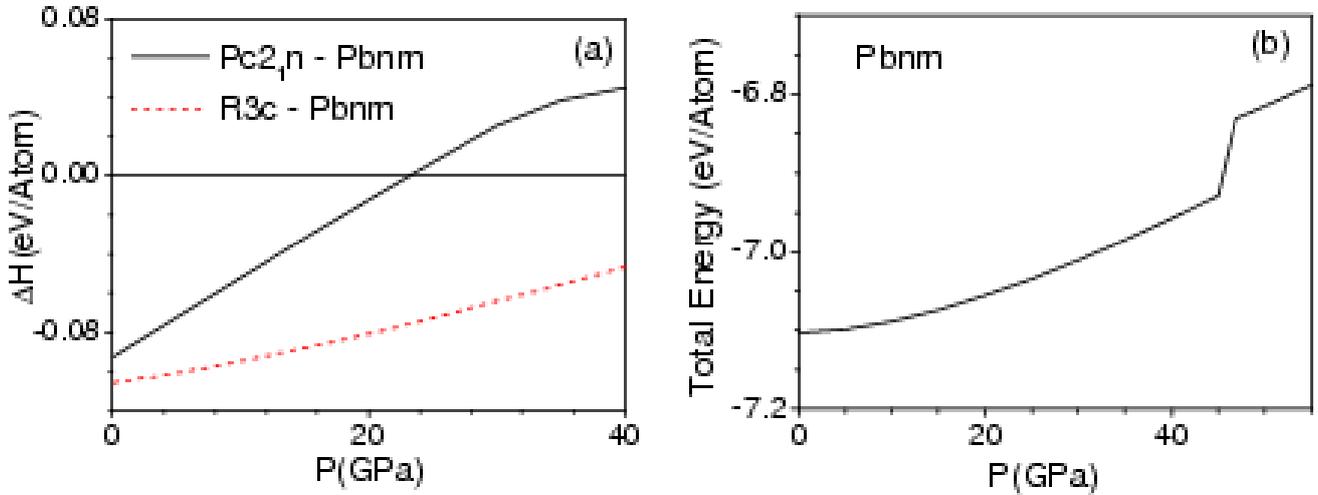

FIG. 8 (Color online) The LDA-calculated equation of state of various phases of GaFeO$_3$ and a comparison with available experimental data [29]. V refers to the volume per formula unit at pressure P. V$_o$ refers to the volume per formula unit of Pc2$_1$n phase at ambient pressure.

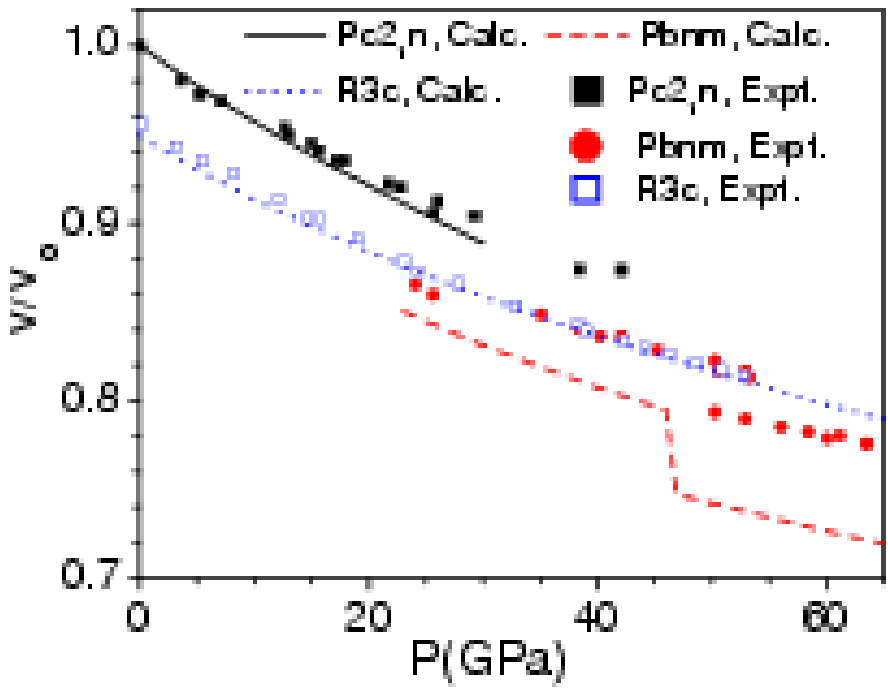

FIG. 9 (Color online) (a) The calculated Grüneisen parameter, $\Gamma(E)$ as a function of energy. (b) The calculated and experimental [28] thermal expansion in the orthorhombic phase (space group $Pc2_1n$).

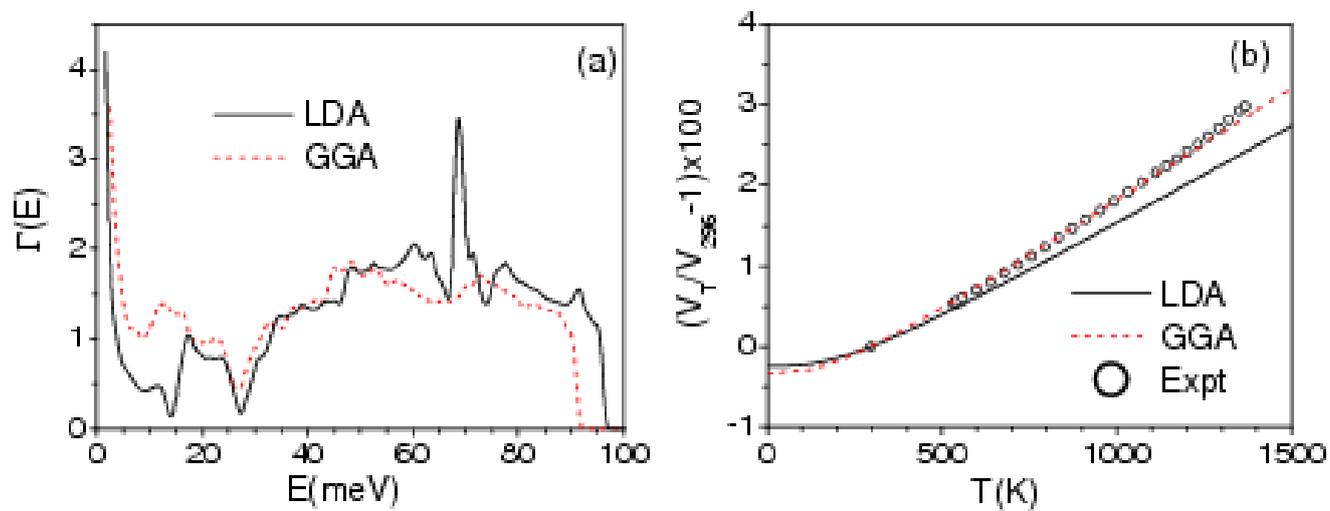